\documentclass[12pt,twoside]{article}
\usepackage{mathrsfs}
\usepackage[margin=1.2in]{geometry} 
\usepackage{amsmath,amsthm,amssymb,mathtools}
\usepackage{commath}

\usepackage{graphicx}
\usepackage{dcolumn}
\usepackage{bm}
\usepackage{authblk}

\graphicspath{ {images/} }
\begin{document}

\title{A unified geometric framework for axisymmetric thin sheet buckling}  

\author[$\dagger$]{Anshuman S. Pal}
\affil[$\dagger$]{James Franck Institute and Department of Physics, University of Chicago, 929 East 57th Street, Chicago, Illinois 60637, USA}
\date{\today}

\begin{abstract}

Thin sheets that are forced at their boundaries develop a variety of shapes aimed at minimising elastic energy of curvature and internal strain. A common approach to characterise these shapes is to impose the condition of isometricity, in which any stretching is confined to a small portion of the sheet. The remainder of the sheet often curves spontaneously in ways that break the symmetry of the sheet and the forcing. That is, the sheet buckles. Characterising this buckling generally requires detailed analysis. Here, we follow a complementary approach, applicable to axisymmetric systems with radial symmetry. It aims to predict the qualitative nature of any buckling using only the simple axisymmetric geometry of the initial system. We view an axisymmetric deformation according to the displacement it imposes on the radial lines making up the sheet. Thus, e.g., in-plane tensile loads on a disk-shaped annulus tend to move radial lines along their length. This deformation amounts to a non-uniform dilation of the annulus. Any compressional strain present in this dilated state then implies symmetry-breaking buckling out of the original plane. Using only these radial motions, we account for the buckling predicted by more detailed analysis: axial wrinkling confined to only the inner part of the annulus. A second type of imposed deformation on radial lines is motion normal to the sheet. Considering the (non-isometric) strains induced by forcing a rotation of these lines through some common angle $\alpha$, one may similarly account for both the buckling of d-cones and that of curved creases. Our framework thus unifies three previously unconnected buckled shapes.


\end{abstract}

\maketitle


\section{\label{sec:intro}Introduction}

Buckling in thin sheets is an archetypical example of elasticity-mediated pattern formation, where the material is forced to displace out of its plane in order to accommodate extra material length. While the traditional method of inducing buckling is by applying forces and torques on the boundary of the sheet, the last decade has seen the emergence of a new paradigm that allows for more fine-tuned loading -- through direct modification of the material's intrinsic metric (i.e. by selectively growing or shrinking the material) \cite{sharon_buckling_2002, muller_conical_2008,klein_shaping_2007,efrati_buckling_2009,sharon_mechanics_2010,  kim_designing_2012,santangelo_buckling_2009}. A key novelty with this second paradigm is that it is explained through the language of (differential) geometry and strains, instead of the traditional language of loads and stresses, and emphasises the geometric origins of buckling. As a result, while the nature of the loading in these two paradigms is very different, they can lead to the same buckled structures, since all material stresses are ultimately related to geometric deformations and strains via constitutive relations \cite{landau_theory_2008}. 

\paragraph{}
While the microscopic origin of buckling is well-known -- due to the presence of compressive strains -- the macroscopic structures that can emerge under buckling show an immense variety, depending on the material moduli, the geometry of the loading, and the presence of external constraints. Recent theoretical work \cite{davidovitch_prototypical_2011} strongly suggests a demarcation between the features determined by energetics (viz. the F\"oppl-von K\'arm\'an equation) and those determined by pure geometry. While energetics, through the balance between stretching and bending, determines the finer structure of the buckling (e.g., the wavelength, wavenumber, etc.), the more large-scale characteristics, and the more global understanding of why and how a structure buckles is purely geometric. This is especially true for \textit{thin} elastic sheets since these are isometric/inextensible, i.e. they obey the strong constraint of length conservation. One way this can be exploited is through Gauss' Theorema Egregium, connecting Gaussian curvature to strain, which implies that there must be a straight line (called a director or generator) passing through every point in the unstrained material \cite{struik_lectures_1988}. The use of directors has been instrumental in explaining the physics of crumpling \cite{lobkovsky_scaling_1995,ben_amar_crumpled_1997, witten_stress_2007}, and the shape of the d-cone \cite{cerda_dcones_1998,cerda_dcones_2005}, the e-cone \cite{muller_conical_2008}, and the narrow curved crease \cite{dias_geometric_2012}. 

\paragraph{}
Besides the isometricity constraint, another thing common to the d-cone and curved crease problems is their \textit{axisymmetric forcing} -- both are formed by applying torques normal to the surface about a circular boundary. A separate class of problems involving axisymmetric \textit{in-plane} forcing, that has been much studied in recent years, is that of azimuthal wrinkling under radial tension (the Lam\'e problem) \cite{geminard_wrinkle_2004,cerda_scars_2005,huang_capillary_2007, pineirua_capillary_2013, vella_capillary_2010, davidovitch_prototypical_2011}. This is the direct equivalent of the problem of tension-induced uniaxial wrinkling \cite{cerda_wrinkling_2003} in a polar geometry. While the small-scale physics of the wrinkling in both cases is determined by energy minimisation, the influence of the curved (circular) boundary in the Lam\'e problem is seen in the large-scale behaviour of the wrinkling -- while uniaxial wrinkling extends everywhere where the applied tension is non-zero, azimuthal wrinkling under radial tension is confined to some critical outer radius, $r_c$, about the inner boundary! Another striking example of the influence of a curved boundary on buckling is seen for the curved crease -- while a flat sheet of paper creased along a straight line stays flat, one creased along a circle (or any closed curve) buckles! 

\paragraph{}
In this paper, inspired by \cite{davidovitch_prototypical_2011} and the geometric approach of Sharon and co-workers, we introduce a unified geometric framework for investigating buckling under axisymmetric (i.e., radial) deformations, that serves to capture the large-scale characteristics of not only Lam\'e wrinkling, but also shapes like the d-cone and the curved crease (where it serves as a complement to the director-based framework). The essence of our framework is to exploit axisymmetry and adopt a \textit{global} (i.e. non-differential) geometric view -- starting with a flat, axisymmetric, strain-free sheet, we consider its radial lines and polar circles to be the basic elastic line elements of our theory. Intuitively, we can thus imagine the sheet to be made up of radial rubber rods with concentric circular hoops attached to them. All radial boundary deformations of this sheet then correspond to deformations of these radial rods -- i) in-plane deformations correspond to radial stretching and squeezing and ii) out-of-plane deformations correspond to rigid rotations -- with the attached circular hoops being constrained to deform as well. 

\paragraph{}
Our full prescription consists of - i) radially deforming the (circular) boundaries, ii) affinely distorting the sheet (i.e. the rods and circles) to satisfy the imposed radial boundary deformations, ii) determining the strain field resulting from the radial displacement field, and iii) identifying the locations and directions of any compressive strain. Then the prediction is that buckling will occur in these locations and along these directions. Using this prescription, and taking into account any extra constraints, we recover the three spontaneous buckling examples mentioned above as different cases of the buckling of the circular hoops. In particular, the advantage of using such a global geometric framework becomes apparent for out-of-plane deformations, since it permits us to `step out' of the plane of the sheet and see the global deformations of the sheet, something not permitted by the local metric. 
\begin{figure}[htb]
\centering
\includegraphics[width=0.8\textwidth]{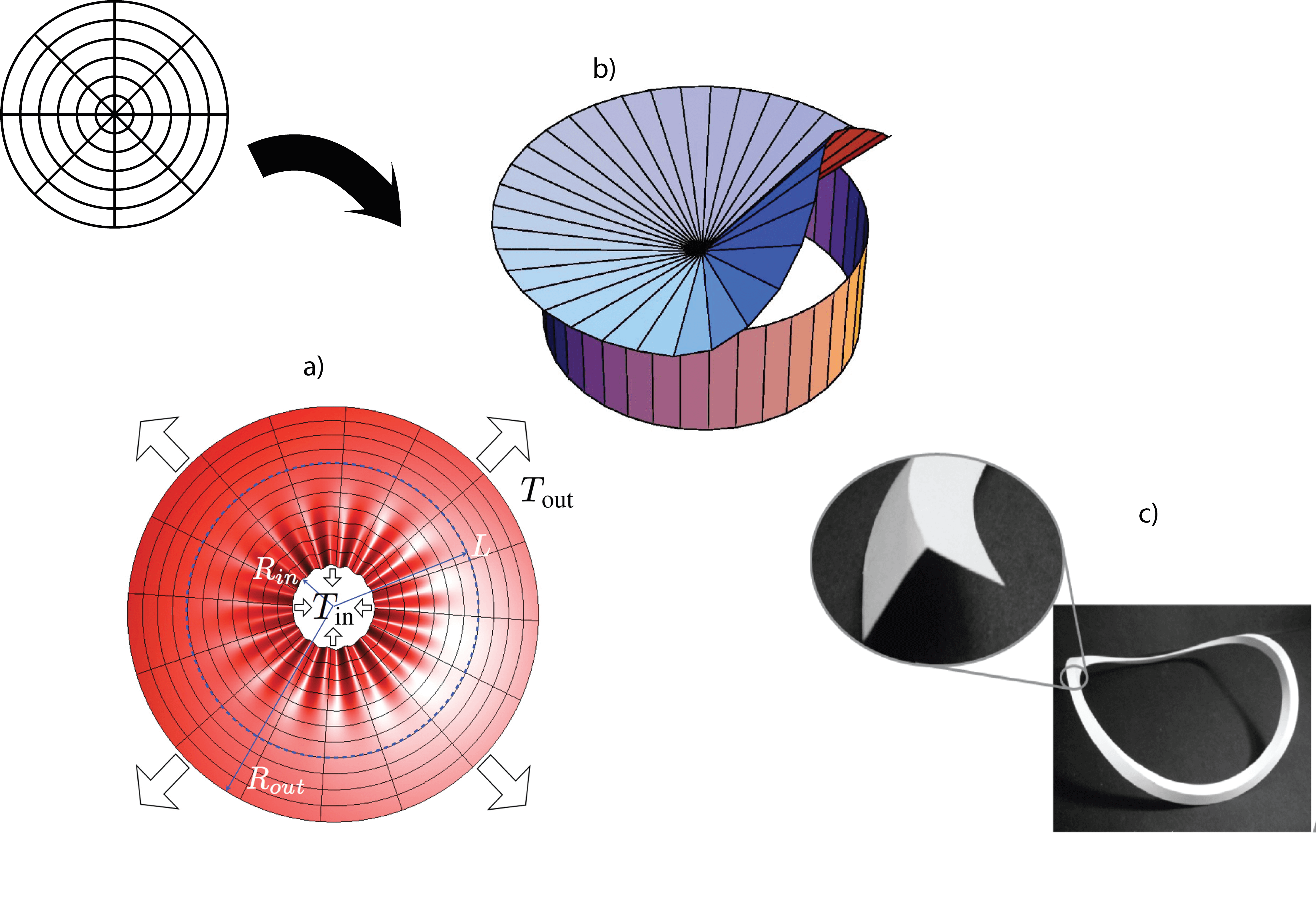}
\caption{The three principal azimuthally buckled structures we cover in this paper - a) Azimuthal wrinkling under radial tension (figure from \cite{davidovitch_prototypical_2011}), b) the d-cone (figure from \cite{cerda_dcones_1998}), c) the curved crease (figure from \cite{dias_non-linear_2014}).}
\label{fig:intro}
\end{figure}
\paragraph{}
The paper is organised as follows. We first introduce our general geometric framework, using the example of in-plane radial deformations. In particular, we focus on the curved line elements (i.e., the concentric circles), since this paper is principally concerned with their buckling (henceforth referred to as azimuthal buckling). Within the category of in-plane deformations -- where the radial lines are stretched lengthwise -- we discuss the Lam\'e problem, and show how our framework can be used to derive a purely geometric upper bound for the critical confining radius $r_c$ in the FT (far from buckling threshold) regime \cite{davidovitch_prototypical_2011} . We then turn to out-of-plane deformations, which involve the radial rotation of the radial lines. Here, we first consider rotations about a fixed pivot -- which may be a point (viz. the origin) or an external circle --  and then rotations about a free pivot. In the former category, among other shapes, we recover the d-cone. In the latter category, we recover the buckled curved crease. In between, we pass through multiple intermediate shapes, which serve to establish a comparison between the d-cone and the curved crease.

\section{General framework: circles and hoop strains} \label{sec:framework}
\begin{figure}[htb]
\centering
\includegraphics[width=0.8\textwidth]{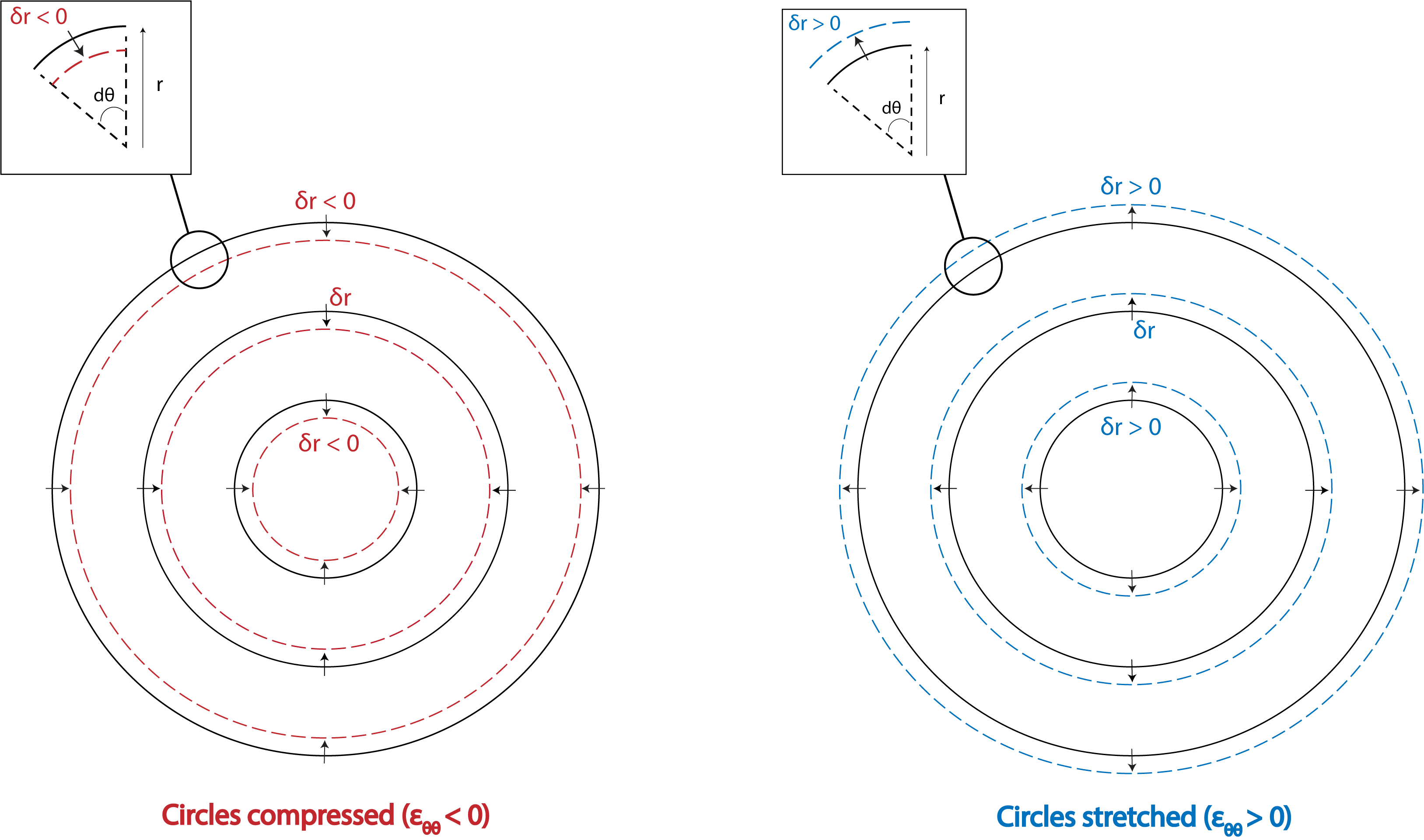}
\caption{Schematic of the example radial deformation described at the beginning of section \ref{sec:framework}. We translate the inner and outer boundaries of an annulus radially inwards by the same distance $\delta r$, so that the radial lengths remain unchanged. However, the attached concentric circles throughout the sheet are (left) contracted for $\delta r<0$, and (right) expanded for $\delta r>0$. Although our framework is global, the insets show that these azimuthal strains are local.}
\label{fig:translations}
\end{figure}

To motivate our framework, we offer an example of the problems we wish to treat in this paper. Consider the deformation of a circular annulus of material by moving both the outer and inner radius inward by a small amount $\delta r$ (Figure \ref{fig:translations}).  In accordance with our prescription, we displace each point in the annulus to accommodate the motion of these boundaries: thus, every point is moved inward by an amount given by the radial displacement field $u_r (r) <0$ (see appendix I for a short proof). This deformation, which may be seen as an inward translation of the radial lines along their length, does not consequently change their net length (i.e., the width of the sheet). However, it does change the circumference of the concentric circles (i.e., the azimuthal distances). Indeed, any material circle of radius $r$ between the outer and inner perimeter has been reduced in circumference by a factor $u_r/r$. Thus, this affinely deformed state has suffered an azimuthal compression everywhere within the sheet. If we now release the requirement of affine deformation and allow the surface to move in the third dimension, it can relieve this compressive strain by means of azimuthal buckling to allow the original circumference of each circle to be restored, thus restoring all distances to their undeformed values. In fact, in any deformation that maintains azimuthal symmetry of the boundaries, we may perform an analogous procedure by inferring the change of circumference in the azimuthal circles in the material.  To this end, we now define the needed quantities for the analysis. 

\paragraph{}
Since elasticity is a theory based on line elements and lengths, our first step is to choose the suitable lines to study in the sheet. Due to the axisymmetry of the deformation, a natural choice is given by the iso-contours of the ($r,\theta$) polar coordinate system -- the mutually orthogonal set of radial lines and concentric circles that span the surface. These lines can be seen as the global, integrated form of the locally-defined axisymmetric metric: 
\begin{equation}\label{eq:metric}
ds^2 = dr^2 + r^2 d\theta^2
\end{equation}
In the following, we show that deviations from the reference/intrinsic metric, which are the cause of elastic strain, can be interpreted as deformations of these global lines. 
\paragraph{}
What sets axisymmetric ($r,\theta$) geometry (or any curvilinear geometry) apart from rectilinear ($x,y$) geometry is its lack of translation invariance; otherwise put, there is a fixed origin. This is evident from the explicit $r$-coordinate dependence of the metric (\ref{eq:metric}). An even clearer way of seeing this is terms of our global line elements and the relation that connects their lengths -- the definition of planar (Euclidean) circles:
\begin{equation}\label{eq:splane}
c_{\rm plane} = 2\pi r.
\end{equation} 
The implication of this equation elastically is that a planar circle of circumference $c_0 = 2\pi r_0$ must live at radial distance $r_0$ from the fixed origin to be unstrained. In other words, the azimuthal and radial directions are constrained\footnote{In fact, the word 'constraint' comes from the Latin \textit{costringere}, which means 'to pull together'}. As we show below, any violation of (\ref{eq:splane}) generates strains. This behaviour stands in stark contrast to rectilinear ($x,y$) geometry, where orthogonal directions are unconstrained by any such geometrical relation as (\ref{eq:splane}); any constraint between the $x$ and $y$ directions is purely material in nature (viz. the Poisson effect). 
\paragraph{}
Starting with this axisymmetric geometry, we apply axisymmetric deformations. In terms of their effects on the line elements, axisymmetry implies certain properties for these deformations. Globally, it means that the radial lines must remain radial, and the circles must remain circles. Locally, it means that a differential arc of length $dc$ at constant radius $r$ from the centre must alway subtend the same angle $d\theta$; in other words, axisymmetric deformations preserve angles (i.e. shape), which means that they don't generate shear strains: $\epsilon_{r \theta} =0$.   

\paragraph{}
We first consider \textit{in-plane} axisymmetric deformations, i.e. where the radial lines are stretched or compressed along their length. Following our prescription, this can lead to one of two possible kinds of violations of the planar constraint (\ref{eq:splane}) –
\begin{enumerate}
\item $c > c_{\rm plane} = 2 \pi r$ (a `hyperbolic violation')
\item $c < c_{\rm plane} = 2 \pi r$ (a `spherical violation') \footnote{The names hyperbolic/spherical refer to the fact that the corresponding circles on a hyperbolic or spherical manifold obey precisely these definitions.}
\end{enumerate}
In the first case, larger circles are forced to live at a radius meant for smaller circles; in the second case, it's the opposite. For the test example considered at the beginning of this section (Figure \ref{fig:translations}), the first case corresponds to radially outward deformations of the circle ($\delta r >0$), and the second corresponds to radially inward ones ($\delta r <0$). For an arbitrary applied boundary deformation, there is a well-defined radial displacement field, $u_r(r)$, that measures by how much a circle in sheet moves inward or outward. As explained at the beginning of the section, such a displacement changes the circumference of the circle at $r$, from $c = 2\pi r \to c' = 2\pi r'$, such that the fractional change in circumference is:
\begin{equation}\label{eq:azimuthalStrain}
\frac{c'-c}{c} =  \frac{2\pi (r'-r)}{2\pi r} = \frac{u_r}{r}, 
\end{equation}
Thus, a radial displacement $u_r$ in circular geometry entails an azimuthal (or 'hoop') strain $\epsilon_{\theta \theta} = u_r/r$, a result familiar from elasticity textbooks (e.g. \cite{landau_theory_2008,gould_introduction_2013}). An outward displacement (case 1.) stretches out the circles ($u_r>0 \implies \epsilon_{\theta \theta} >0$) , and an inward displacement (case 2.) squeezes in the circles ($u_r <0 \implies \epsilon_{\theta \theta} <0$). The elastic response to such strain depends on the sign -- while negative strain (excess length) can be relieved relatively easily, by buckling out of the plane, positive strain (insufficient length) is generally difficult to relieve (an example of such strain relief is fracture). The concept of hoop strain as a change of circular circumference is essential to the rest of this paper, and the different cases of azimuthal buckling considered are all shown to be consequences of \textit{negative} hoop strain.

\paragraph{}
 We illustrate our framework using a straightforward example, where the axisymmetric strains are generated not by radial deformations, as in the rest of the paper, but by direct modification of the reference/material metric. Consider the case where the circumference $c$ of the circles is modified, at fixed radius $r$, thereby generating only azimuthal strain. If this change scales linearly with $r$, it is known as a disclination or angular defect \cite{kleman_disclinations_2008} -- since it involves changing the angular extent of the arc from $2\pi \to 2 \pi + \Phi$ ($\Phi$ is called the angular excess) -- and gives rise to conical solutions (c.f. Section \ref{sec:rotations}). Physically, such defects can be seen as a consequence of pure azimuthal growth ($\Phi >0$) or shrinkage ($\Phi <0$) \cite{muller_conical_2008}. They also correspond to our criterion of axisymmetry since the initial circles remain circles and the radial lines remain radial. Now, for $\Phi >0$, we have a hyperbolic violation of (\ref{eq:splane}) since $c > c_{\rm plane}$, which means that a larger circle is being forced to fit into a smaller space. The fractional excess length is then given by 
\begin{equation}\label{eq:disclination}
\frac{c-c_{\rm plane}}{c_{\rm plane}} =  \frac{(2\pi +\Phi)r - 2\pi r}{2\pi r} = \frac{\Phi}{2\pi}. 
\end{equation}
The azimuthal strain, however, is given by the negative of this: $\epsilon_{\theta \theta} = -\frac{\Phi}{2\pi}< 0$, since the strain measures the (fractional) deviation of lengths from the reference metric (i.e. the unstretched lengths of the sheet). For the circles here, the unstretched length is $c$, and $c_{\rm plane}$ acts as the deviation. Thus, as before with the deformations, the circles feel negative strain, which allows them to buckle; the resultant shape is called an e-cone. If $\Phi <0$, then $\epsilon_{\theta \theta} >0$, and the circles cannot buckle; but they can relieve this azimuthal tension by axisymmetrically rotating the radial lines, and turning into a regular cone (also called a c-cone). Such conical solutions play an integral role later in Section \ref{sec:rotations},


\section{Radial in-plane deformations: finite extent of azimuthal wrinkling} \label{sec:displacements}
The first category of deformations we discuss are radial \textit{in-plane} deformations -- in which the circles are strained according to (\ref{eq:azimuthalStrain}). These can be implemented experimentally via a radial metric change \cite{mora_buckling_2006}, or more commonly, by subjecting a circular annulus to radial forces. In particular, the problem of azimuthal wrinkling under radial tensile loads (the Lam\'e problem) has recently been widely studied \cite{davidovitch_prototypical_2011, geminard_wrinkle_2004,cerda_scars_2005,huang_capillary_2007,vella_capillary_2010}, leading to experimentally confirmed scaling laws for the wavelength and amplitude. However, unlike rectilinear (uniaxial) wrinkling \cite{cerda_wrinkling_2003}, which is decribed entirely via its wavelength and amplitude, azimuthal wrinkling has another associated length scale – $r_c$, a critical outer radius marking the extent of wrinkling (cf. Figure \ref{fig:intro}a ). Moroever, unlike unaxial wrinkling, azimuthal wrinkling is caused by a pull at the inner boundary, and opposed by a pull at the outer boundary, with a non-trivial value of $r_c$ existing only in the presence of pulls both at the inner and outer boundaries. A complete analysis of the wrinkling, including expressions for $r_c$, has been given in \cite{davidovitch_prototypical_2011}, using a energy minimisation approach. But the \textit{existence} of the finite confining radius $r_c$ can be predicted as a pure consequence of the curved boundaries of the annulus and of the sign of the displacement loads at the two boundaries. In the following, we use our framework to derive upper bounds for $r_c$ in both the NT (near buckling threshold) and FT (far from buckling threshold) regimes, highlighting the origin of the different scalings in the two regimes. Our analysis thus complements that given in \cite{davidovitch_prototypical_2011} and elsewhere \cite{geminard_wrinkle_2004,pineirua_capillary_2013}.

\paragraph{}
For this, we need to consider the sign of $\epsilon_{\theta \theta} = u_r/r$ under the action of different boundary deformations. Whereas radial tensile deformations always stretch the radial lines, their action on the circles is very different. If there is only a pull at the inner radius, this means that all the circles in the annulus are pulled inward ($u_r < 0$ for all $r$) and compressed, creating a tendency to buckle. Conversely, if there is only a pull at the outer radius, all the circles are pulled out ($u_r>0$ for all $r$), and thus the whole sheet is in tension. However, for a combination of these two pulls, provided the outer pull is strong enough, we can have a switch in the sign of $u_r$! More concretely, if $\Delta_{\rm in }$ and $\Delta_{\rm out}$ be the displacement loads at the inner and outer radii $R_{\rm in}$ and $R_{\rm out}$ resp., then $u_r$ must obey the boundary conditions $u_r(r_{\rm in}) = \Delta_{\rm in}$ and $u_r(r_{\rm out}) = \Delta_{\rm out}$. Thus, if  $\Delta_{\rm in} < 0$ and $\Delta_{\rm out}>0$ (both pulls), then we must have $u_r =0 $ at some intermediate radius (see appendix). By our arguments, there cannot be any buckling beyond this intermediate radius, and any observed azimuthal buckling must be confined to the inside of this radius corresponding to $u_r = 0$. This clearly shows the existence of a finite confining radius, $R_{\rm in} < r_c < R_{\rm out}$. 
\paragraph{}
The actual value depends on the form of $u_r$, which cannot be given simply by geometric reasoning. This is because the azimuthal strains generated by $u_r$ in turn generate radial strain, that then modify the $u_r$ field -- the same reason that the deformation $\Delta_{\rm in} = \Delta_{\rm out}=\Delta$ does not give a trivial $u_r =\rm {const.} =\Delta$ field  As a result, a self-consistent $u_r$ can only be determined through energy minimisation. The upper bound obtained from this $u_r$ field (see appendix) gives the correct NT scaling: $r_c^{NT} \sim \sqrt{\frac{\Delta_{\rm in}}{\Delta_{\rm out}}}$, where $\frac{\Delta_{\rm in}}{\Delta_{\rm out}}$ is the appropriate 'confinement parameter' in our problem (cf \cite{davidovitch_prototypical_2011}). On the other hand, the FT regime (appropriate for thin, highly bendable sheets) is where the compressed azimuthal circles have fully buckled ($\epsilon_{\theta \theta}\to 0$), leaving the radial lines unconstrained. Thus, in the region of buckling ($R_{\rm in} <r \leq r_c$), one no longer needs to use energetics to find the zero of $u_r$. If we presume this azimuthal strain collapse ansatz to exist upto the outer radius, then we can derive a purely geometric outer bound for $r_c$. In the limit, $r_c \to R_{\rm out}$ (i.e. for $\Delta_{\rm out} \to 0^+$), the bound becomes exact.
\paragraph{}
In this limit, the radial strain generated by the displacement loads $\Delta_{\rm in }$ and $\Delta_{\rm out} $ is constant. Thus, we get:
\begin{equation}\label{eq:ur_translations}
u_r (r) = \Delta_{\rm in } + \frac{\Delta_{\rm out} - \Delta_{\rm in}}{R_{\rm out}-R_{\rm in}} (r-R_{\rm in}) ,
\end{equation}
with $\Delta_{\rm in},\Delta_{\rm out} >0$. The radius where this crosses zero gives us the said upper bound for $r_c$:
\begin{equation} \label{eq:rcBound}
r_c^{FT} \leq R_{\rm in} + \frac{\Delta_{\rm in}}{\Delta_{\rm in} - \Delta_{\rm out}}(R_{\rm out} - R_{\rm in}).
\end{equation}
If, like \cite{davidovitch_prototypical_2011}, we consider $\Delta_{\rm out} >0$ fixed, and apply $\Delta_{\rm in} <0$, then expanding (\ref{eq:rcBound}) to lowest order in $\Delta_{\rm in}/\Delta_{\rm out} $, we get 
\begin{equation}\label{eq:rc_linear}
r_c^{FT} \leq R_{\rm in} +  \abs{\frac{\Delta_{\rm in}}{\Delta_{\rm out}}} (R_{\rm out} - R_{\rm in})  + \mathcal{O}(\frac{\Delta_{\rm in}}{\Delta_{\rm out}}^2) ,
\end{equation}
which gives us back the expected linear FT scaling: $r_c^{FT} \sim\Delta_{\rm in}/\Delta_{\rm out} $. Notably, unlike \cite{davidovitch_prototypical_2011}, this derivation also recognises the existence of the outer boundary at $R_{\rm out}$, and gives the correct limit $r_c \to R_{\rm out}$ for $\Delta_{\rm out} \to 0$, as expected since it saturates the bound and accurately represents the physics at this limit. As a result, an expansion of (\ref{eq:rcBound}) around this opposite limit, of $\Delta_{\rm in} <0$ fixed and small $\Delta_{\rm out}>0$, should give the correct value of $r_c$ as well, something that cannot be obtained from the treatment of \cite{davidovitch_prototypical_2011}:
\begin{equation}
r_c^{FT} = R_{\rm out} - \abs{\frac{\Delta_{\rm out}}{\Delta_{\rm in}}} (R_{\rm out}-R_{\rm in}) + \mathcal{O} (\frac{\Delta_{\rm out}}{\Delta_{\rm in}}^2) .
\end{equation}
Unfortunately, this is precisely the regime not considered in the experiments cited above, and so, there is little data available to which to compare this result.
 
\paragraph{}
Finally, to close this section, we go further and consider an exhaustive classification of possible buckling behaviours of annular sheets under radial displacements, depending on the signs of $\Delta_{\rm in}$ and $\Delta_{\rm out}$. To do so, we consider the simplest possible cases where a deformation is applied on only one boundary, and the other boundary is held fixed; since we only consider the signs, there are only four such categories. The usefulness of this lies in that any arbitrary deformation imposed on both boundaries can be constructed as a linear combination of these four cases:

\begin{enumerate}
\item	$\Delta_{\rm out} < 0 $ (compressive), $\Delta_{\rm in} = 0$
\item	$\Delta_{\rm out} > 0 $ (tensile),  $\Delta_{\rm in} = 0$
\item	$\Delta_{\rm in} < 0 $ (tensile),  $\Delta_{\rm out} = 0$
\item	$\Delta_{\rm in} > 0 $ (compressive),  $\Delta_{\rm out} = 0$
\end{enumerate}

In terms of azimuthal strains, however, cases 1. and 3. and cases. 2. and 4. are in fact degenerate. As can be seen from (\ref{eq:ur_translations}), 1. and 3. create $u_r <0$ everywhere in the sheet and cause wrinkling, while 2. and 4. create $u_r >0$ and oppose wrinkling. More interesting physics occurs when we impose \textit{both} non-zero $\Delta_{\rm in}$ and $\Delta_{\rm out}$, with opposite signs, to combine the above opposing tendencies. Thus, a combination of 2. and 3. produces radial tension and azimuthal strain that changes sign -- giving rise to finite-range azimuthal wrinkling.  But what happens for a combination of 1. and 4.? These deformations generates radial compression and azimuthal tension (since $u_r >0$ everywhere), and should thus give rise to axisymmetric, radially buckled shapes. To the authors' best knowledge, such buckling has not been investigated before, and is the subject of current work.

\section{Radial out-of-plane deformations} \label{sec:rotations}

Having considered the effects of axisymmetric in-plane displacements, which produce both radial and azimuthal strains, we now consider the effects of axisymmetric out-of-plane deformations. This means that we choose a particular circle on the sheet $r=r_p$ (or the centre, $r_p =0$) as a \textit{fixed} pivot about which we rotate the radial lines, either inwards towards the central axis, or outwards out from it. Such a rotation then corresponds to making a 'conical' shape out of the annulus or disk. 

\paragraph{}
Besides necessitating the use of three-dimensional cylindrical coordinates, there is one crucial difference between radial in-plane displacements and radial out-of-plane rotations in terms of the strains they generate. Since the radial lines are simply rotated about, they are left unstrained -- $u_r=0$ -- and all in-plane radial distances are untouched. However, these rotations do strain the attached circles, and consequently, all compression and buckling under radial rotations is azimuthal. Moreover, the absence of any transverse, radial tension has a profound effect on the buckled shape of the circles.  Since this buckling is unconstrained, i.e., there is no penalty for large-wavelength, large-amplitude buckling \cite{cerda_wrinkling_2003, paulsen_curvature-induced_2016}, we should not expect the circles to wrinkle -- a buckling solution that costs high bending energy. Indeed, in the shapes discussed below, in the absence of other external constraints, we find that the circles always buckle spontaneously from the planar state to the minimum possible wave-number state.

\paragraph{}
This section is arranged as follows. We first calculate the strains generated by an arbitrary radial rotation, and then separately consider the cases of rotation about the centre ($r_p  =0$), and rotation about an external circle ($r_p \neq 0$). Here, and in the rest of the paper, we try to make what conclusions we can about the buckled shapes, using purely geometry, length conservation, and the role of boundary conditions and topology; such a purely geometric treatment gives the exact results for physical sheets in the zero thickness limit. In particular, we give considerable attention to the point rotation ($r_p  =0$) case and its consequent conical solutions -- the buckled d-cone and the planar overlapping c-cone -- since they obey many unique properties that can be shown using purely geometry. They thus serve as ideal reference shapes for comparing the other radial rotation-generated shapes.

\subsection{Calculating strains}
Following our framework, we start with a planar, axisymmetric system, apply an axisymmetric boundary deformation, displace each point in the annulus to accommodate the motion of the boundaries, and see what strains develop. We work with a planar annulus of inner and outer radii $R_{\rm in}$ and $R_{\rm out}$, but the results extend to disks as well ($R_{\rm in } =0$). Since we are now considering the third dimension, we henceforth work with cylindrical coordinates ($\rho, \phi, z$). $r$ now denotes the radial distance from the centre of axisymmetry, and $d\theta$ denotes the angle subtended by an arc at this centre. Of course, in the pre-rotation flat state, $\rho$ and $r$ coincide. Thus, the same differential arc length can be written as $dc = r d\theta= \rho d\phi$. Also, the definition of circles (\ref{eq:splane}) must be re-written as:
\begin{equation}
c_{\rm plane} = 2\pi \rho.
\end{equation}
\begin{figure}[htb]
\centering
\includegraphics[width=0.9\textwidth]{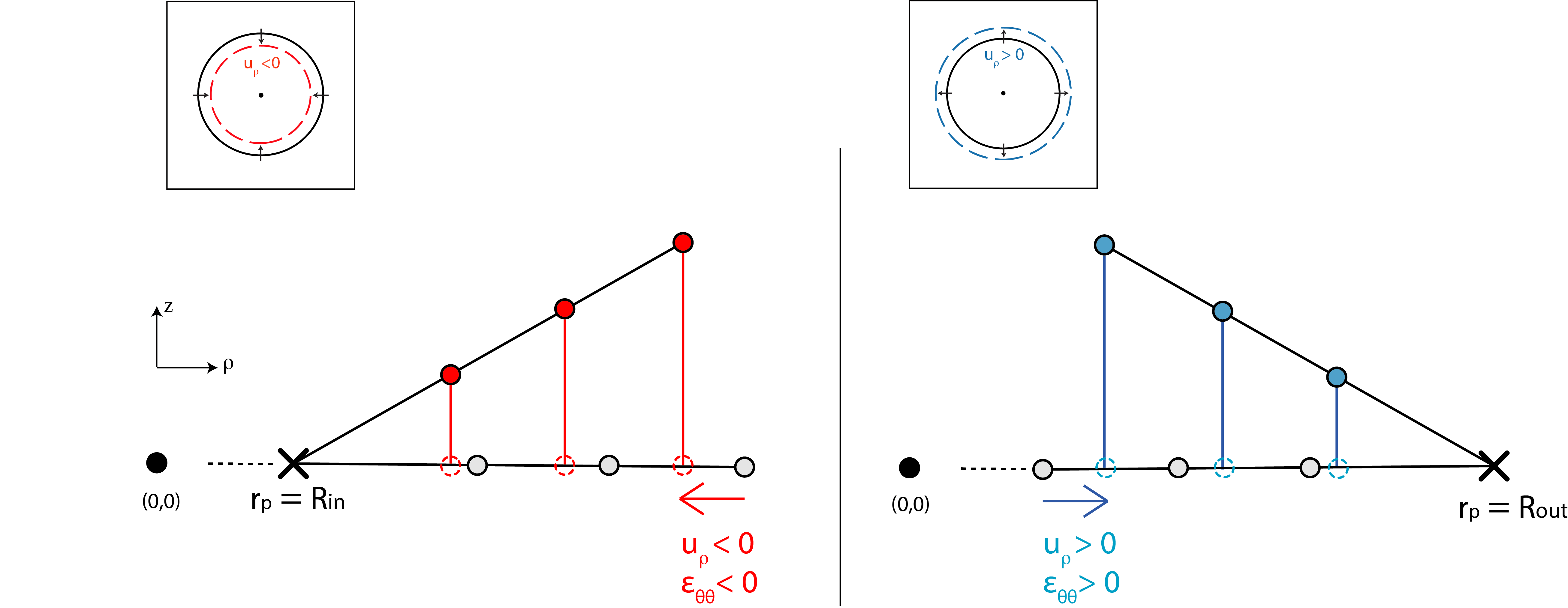}
\caption{A 1D schematic of the hoop strains developed in the circles for the two limiting cases of rotations about a) about the inner radius $R_{\rm in}$, and b) about the outer radius $R_{\rm out}$. Red marks compression ($\epsilon_{\theta \theta} <0$) and blue marks tension ($\epsilon_{\theta \theta} >0$).}
\label{fig:rotationLimits}
\end{figure}

Consider a radial rotation of angle $\alpha$ ($0 < \alpha \leq \pi/2$). There are two limiting cases of these deformations --  i) inward rotation about $R_{\rm in}$, and ii) outward rotation about $R_{\rm out}$ (Figure \ref{fig:rotationLimits}).  In i), all the circles -- except the pivot circle, of course -- are moved inwards ($u_\rho <0$) and thereby compressed ($\epsilon_{\theta \theta} <0$). In ii), they are all stretched ($\epsilon_{\theta \theta} >0$). Following this logic, a rotation about an arbitrary pivot circle $R_{\rm in} < \rho_p < R_{\rm out} $ produces azimuthal tension on the inside, and compression on the outside. We get the following expressions for the displacement fields:
\begin{align}
u_\theta &= 0 \nonumber \\ 
u_\rho &= (r-r_p)(\cos \alpha -1) =  (\rho-\rho_p)(\cos \alpha -1)\\
u_z &= (r-r_p) \sin \alpha =  (\rho - \rho_p) \sin \alpha. \nonumber
\end{align}
Thus, the only non-zero strain is the azimuthal strain, given by the fractional change in circumference of the circles : 
\begin{equation}\label{eq:rotationClampedStrain}
\epsilon_{\theta \theta} (\rho) = \frac{u_\rho}{\rho} = (1-\frac{\rho_p}{\rho})(\cos\alpha -1).
\end{equation}

\subsection{$\rho_p = 0$: conical solutions }

\begin{figure}[htb]
\centering
\includegraphics[width=0.9\textwidth]{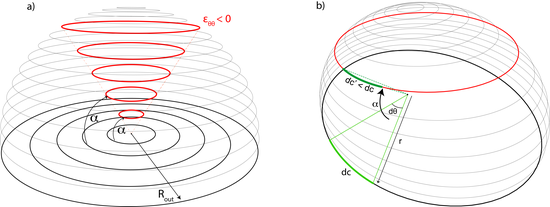}
\caption{A rotation of angle $\alpha$ about the origin ($\rho_p =0$) corresponds to making a cone out of a planar sheet (here, a disk). (Left) The circles in the sheet are compressed into the $\alpha$-latitudes of their respective spheres. Each circle is equally compressed: $\epsilon_{\theta \theta} = \cos \alpha-1$ (red signifies compression). (Right) Focusing on the outer sphere, we see the local picture of the same defomation, with each arc $dc$ (highlighted in green) of the circle being squezzed in a smaller length $dc' = dc \cos \alpha$.}
\label{fig:cones}
\end{figure}

We start by seeing why the above-defined radial rotation strains the circles. First, consider the effect of the above-defined rotation, of angle $\alpha$ about the origin ($r_p =0$), on a differential arc element of length $dc$ located on the equator at the planar coordinates ($r, \theta$). If we consider ($\hat e_r, \hat e_\theta$) to be the local planar basis, then the radial rotation corresponds to a rotation of $dc$ by an angle $\alpha$ about the direction $\hat e_\theta$; this preserves the length $dc$. However, if we consider the entire equator at radius $r$, which is made up of a continuum of such differential arc elements, then each of these elements, located at a different polar angle $\theta$, is rotated about a different axis $\hat e_\theta$. Such a transformation need not preserve lengths. Indeed, as seen in Figure \ref{fig:cones}, the result of this transformation is that the equator is rotated to the latitude $\alpha$ of the sphere of radius $r$. Since there is less circumference available at latitude $\alpha$, it means that the circle is compressed ($\epsilon_{\theta \theta} <0$). More precisely, since the distance to the central axis is now $\rho= r\cos\alpha < r$, it means that the fractional change in circumference as $c = 2\pi r\to c'=2\pi \rho$ is
\begin{equation}\label{eq:coneStrain}
c'-c = \frac{\rho -r }{ r} = \cos \alpha-1 <0.
\end{equation}
This is, of course, the value of $\epsilon_{\theta \theta}$ obtained by putting $\rho_p = 0$ in (\ref{eq:rotationClampedStrain}) above. Thus, locally, each differential arc of length $dc$ is now squeezed into a length $dc \cos \alpha$. 
\paragraph{}
The fact that the compression is the same for all circles -- $\epsilon_{\theta \theta} = \cos\alpha -1 =\rm const.$ --  means that the $\rho_p = 0$ rotation is independent of any system scale, and hence, so must be any buckled structure it leads to. Such a solution can be termed a 'cone', since the scale-independent buckling of the circles is equivalent to saying that the straight radial lines act as surface directors/generators and converge at a point, namely the tip of the cone (c.f. \cite{cerda_dcones_1998, witten_stress_2007, muller_conical_2008}). Equivalently, the scale-invariance means that a conical solution is characterised by a constant (dimensionless) angular extent, which acts as a point Gaussian charge for the solution.

\paragraph{}
This scale-invariance has a couple of notable consequences for the buckled solution (see the appendix for justifications). Firstly, it implies that all the circles buckle similarly, with the circle at initial radius $r$ living on the sphere of radius $r$ in its buckled state. Secondly, the presence of a radial director passing through each point in the sheet means that the Gaussian curvature is zero, i.e., the surface is developable. However, it does not assure that the radial directors are unstrained along their length. Indeed, it can be shown that the conical solution is the only radially rotated buckled solution that keeps the radial lines unstrained, and is thus fully strain-free: $\epsilon_{rr} = \epsilon_{\theta \theta}=0$ \cite{ben_amar_crumpled_1997}. As a corollary, a non-conical solution formed under radial rotations -- i.e. a buckled shape where the circles do not live on a sphere, or equivalently, where the directors are not radial -- must necessarily be strained. This statement will be of use to us later. 

\paragraph{}
We have already seen two examples of cones at the end of Section \ref{sec:framework} -- the e-cone and the c-cone -- which are produced by the in-plane strains generated by selective growth or shrinkage of the concentric circles (i.e. by direct metric modification). We now discuss the two possible conical solutions that can be produced via a radial rotation of angle $\alpha$ -- the d-cone and the overlapping c-cone. The former highlights the role of confining boundary conditions in permitting a buckled solution, while the latter shows how changing the topology permits a simpler, planar solution. Both together later serve as a reference basis for comparing the curved crease and its cut equivalent.  

\subsubsection{The d-cone: the buckled solution}
As stated above, under a conical rotation, all the circles buckle similarly; thus, it suffices to consider any one circle, and any derived conclusions extend equally to all of them, i.e., to the entire sheet. Now, whether the compressed circle can buckle in order to relieve its compression depends on the rigidity of the imposed boundary condition (BC). Denoting by $\alpha (\theta)$ the local latitude along the circle, if we impose $\alpha (\theta) =\alpha_c$ (i.e. force the annulus to live exactly on a cone of angle $\alpha_c$), then buckling will be impossible (see Figure \ref{fig:cones}). However, if we allow the annulus to move freely inside the cone, i.e. if we impose only $\alpha (\theta) \geq \alpha_c$, then there will be enough freedom for it to buckle and form a non-trivial structure -- what is called the developable cone or \textit{d-cone} \cite{witten_stress_2007,cerda_dcones_1998,cerda_dcones_2005}. $\alpha (\theta) \geq \alpha_c$ is precisely the BC applied in the classic d-cone experiment \cite{cerda_dcones_1998}, which involves forcing a sheet via a normal point force into a supporting ring radius $R$  through a distance $\delta$, giving us $\alpha_c = \tan^{-1}(\delta/R)$ (Figures \ref{fig:cones} and \ref{fig:intro}b). We consider this to be equivalent to a disk being confined inside a cone of angle $\alpha_c$. 
\paragraph{}
The final shape of the d-cone will thus be given by the shape of a single buckled circle at arbitrary radius $r$ -- given by the minimal bending energy curve of length $2\pi r$ living on the $r$-sphere, with its latitude restricted to $\alpha (\theta) \geq \alpha_c$. The solution to this problem can be found in \cite{cerda_dcones_1998,cerda_dcones_2005,cerda_elements_2004}, and shows that the energy minimisation leads to an Elastica equation with periodic BCs. Thus, the whole cone is described by the buckling of a single circular rod. 

\subsubsection{The overlapping c-cone: the planar solution}
There is a simpler, trivial way of getting rid of the azimuthal compression, $\epsilon_{\theta \theta} = \cos\alpha -1$. Instead of relaxing the boundary condition, $\alpha(\theta) =\alpha_c$, we change the topology of the annulus, viz. we cut all the circles by cutting up the annulus radially (henceforth called a 'cut annulus'). In this way, all the excess length can be lost simply by overlapping, instead of buckling, and the resultant cone is described by an overlapping planar circle, lying on the $\alpha_c$-latitude of its sphere. Since the circles are planar and unbuckled, we recognise this shape as nothing but an overlapping version of the c-cone described at the end of section \ref{sec:framework}, made via an azimuthal metric modification. Moreover, as expected for a cone (c.f. (\ref{eq:disclination})), the overlapping c-cone can be shown to have constant angular overlap: $\Phi = 2\pi \cos \alpha_c$. 

\subsubsection{Properties of conical solutions}
For future reference, we summarise the special properties of conical solutions here:
\begin{enumerate}
\item Conical solutions are the only fully strain-free solutions possible under out-of-plane rotations. In the absence of additional constraints, they thus form spontaneously under out-of-plane deformations.
\item They are scale-invariant solutions. Thus, they are insensitive to any material length scales ($R_{\rm in}, R_{\rm out}$).
\item Because of scale-invariance, there is no intrinsic length scale associated to the strain-relief mechanisms (buckling and overlapping). Thus, they are low-wavelength in nature, making them sensitive to the global topology of the sheet. Any higher-wavelength, local buckling requires some additional constraint that generates an external length scale, like clamping or gravity (see below).
\item A two-dimensional cone can be entirely described in terms of a one-dimensional curve, which moreover lies on a sphere. In our framework, the curve on the $r$-sphere coincides with the circle of radius $r$, while the radial lines act as the generators. 
\end{enumerate}

\subsection{$\rho_p \neq 0$: clamped rotations about circles}

\begin{figure}[htb]
\centering
\includegraphics[width=\textwidth]{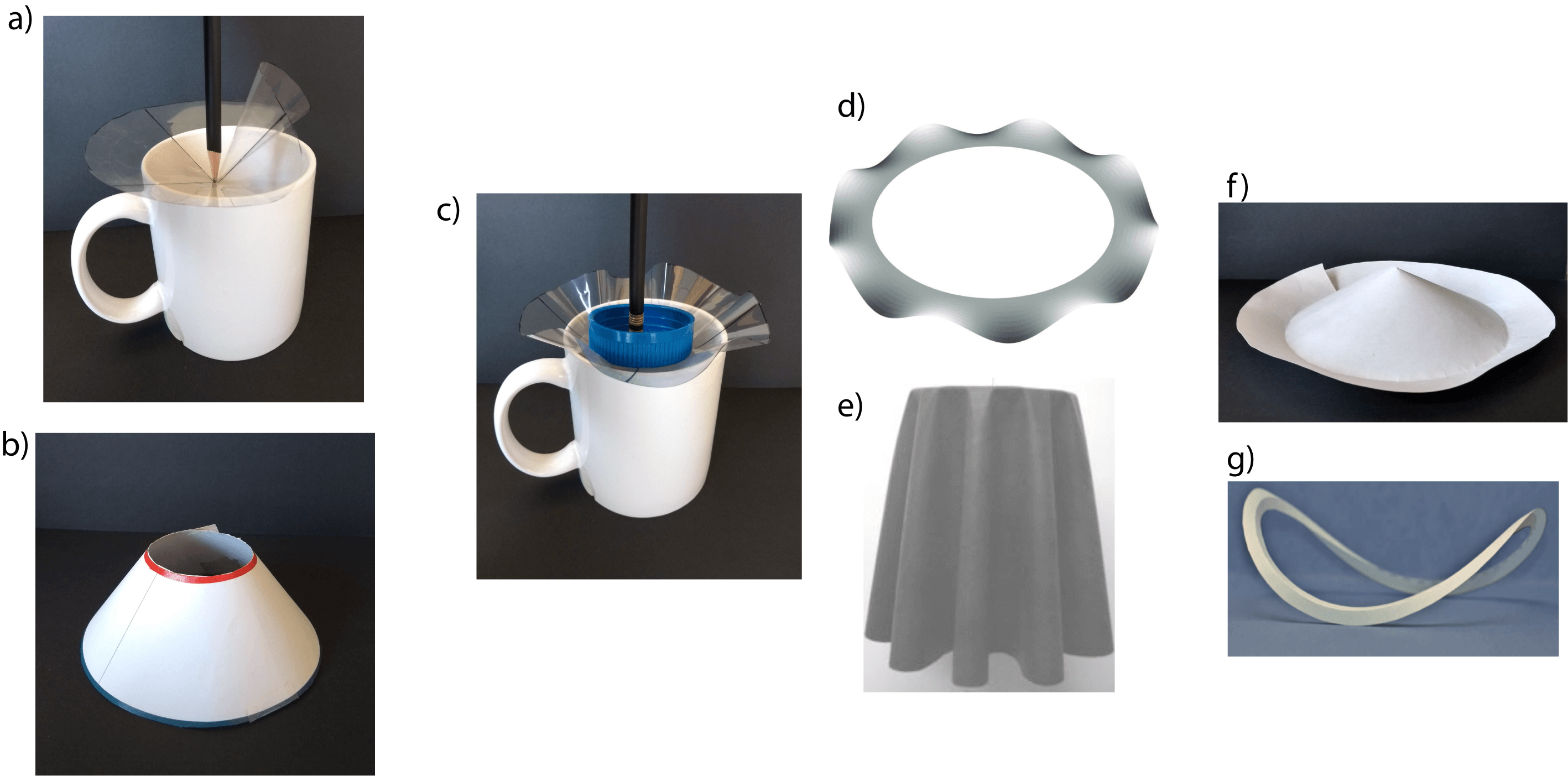}
\caption{Examples of buckled and non-buckled shapes produced via radial rotations. a-b) are simple conical shapes produced by point rotation ($\rho_p =0$), c-e) show azimuthal ruffling produced by clamped rotations about circles ($\rho_p \neq 0$), and f-g) are circularly creased shapes. a) A d-cone formed by point indentation of a transparency disk; b) an overlapping c-cone formed by external rotation about an unfixed pivot (c.f. Section \ref{sec:creases}). c) The d-cone experiment repeated with a finite-radius object leads to azimuthal ruffling; d) (taken from \cite{mora_buckling_2006}) shows a similar pattern in a Lam\'e experiment with radial in-plane tension and clamped BC applied at the inner boundary ; e) (taken from \cite{cerda_elements_2004}) shows a large, rubber sheet pulled down by its own weight, in an experiment similar to c), but with gravity providing a new wavelength scale. f) A circular crease in a radially cut disk leads to a shape where the inside and outside surfaces are both conical (a 'bi-cone'); g) in an uncut annulus, the same crease generates buckling, but the resultant shape is not a double d-cone (figure taken from \cite{dias_geometric_2012}).} 
\label{fig:rotationShapes}
\end{figure}

From (\ref{eq:rotationClampedStrain}), we see that in contrast to the $\rho_p =0$ (d-cone) case discussed above, the $\rho_p \neq 0$ case generates both compressive and tensile azimuthal strains -- $\epsilon_{\theta \theta} <0$ on the outside, and $\epsilon_{\theta \theta} >0$ on the inside. While compressive strains can be easily relieved through buckling, tensile strains can usually not be relieved through any mechanism short of fracture; as a result, any deformation that generates tensile strains is difficult to realise. One way, however, to physically realise $\rho_p \neq 0$ rotations is by clamping down the surface on one side of $\rho_p$, and rotating the unclamped side about the edge of the clamp. In this way, we can isolate and study the response of the unclamped side. 

\paragraph{}
We first consider the case where the exterior is rotated inwards, since this leads to compression and buckling. One way of doing this deformation is via a modified form of the classic d-cone experiment, where the point object is replaced by a disk of radius $\rho_p$. Figure \ref{fig:rotationShapes}c shows such an experiment on a circular transparency; we see that the buckling takes the form of local ruffling (the rotational equivalent of wrinkling). Since this ruffling is local, the wavelength scale $\lambda$ must be set by some local length scale. Here, the only such length scale available is $w$, the width of the exposed annulus, giving us $\lambda \sim w$. Note that the number of wrinkles $m$ is constant, which means that all the circles buckle similarly; once the wavelength scale is set, the wavelength must increase linearly with distance from $\rho_p$ since the length of each circle increases linearly as well. As for the conical solutions, then, this indicates that the rotated radial lines are directors of the surface.

\paragraph{}
Similar wrinkling patterns, with similar or different loads, can also be found in the literature in different contexts. \cite{mora_buckling_2006} discovered a similar geometric azimuthal ruffling with $\lambda \sim w$, while studying discontinous radial growth in gels. This problem was shown to be equivalent to the Lam\'e problem with clamped inner boundary and only an inward tension. While this Lam\'e wrinkling is induced by in-plane displacements and is thus purely along $z$, our wrinkling is generated by rotations and has both $\rho$ and $z$ components, and thus looks very different; however, our geometrical reasoning using circles shows that their underlying structure is the same. Another example, this time involving rotational deformations, can be seen in the draping patterns when a cloth is draped over a circular table \cite{cerda_elements_2004}. The table here provides the clamping, while gravity provides the rotational torque. As expected, this gives rise to azimuthal wrinkling, but with the wavelength scale now set by a combination of the sheet's material properties and its weight.
\paragraph{}
In the inverse case (only possible with an annulus), the exterior of $\rho_p$ is clamped and the interior is tried to be rotated. However, since such a deformation generates $\epsilon_{\theta \theta} >0$, it is almost impossible to perform. A strong pull, as expected, leads to a radial tear.

\section{Rotations about unfixed pivots: curved creases} \label{sec:creases}
We now consider the class of rotations where the pivot circle is no longer held fixed, but is allowed to translate. These deformations are thus combinations of the in-plane displacements and out-of-plane rotations considered till now. Moreover, with curved creases in mind, we also extend the 'mono-rotation' (single rotation angle $\alpha$) considered before to a 'bi-rotation' -- i.e., a deformation where the two sides of the pivot circle $\rho_p \neq 0$ are rotated by two different angles $\alpha_{\rm in}$ and $\alpha_{\rm out} $.

\begin{figure}[htb]
\centering
\includegraphics[width=\textwidth]{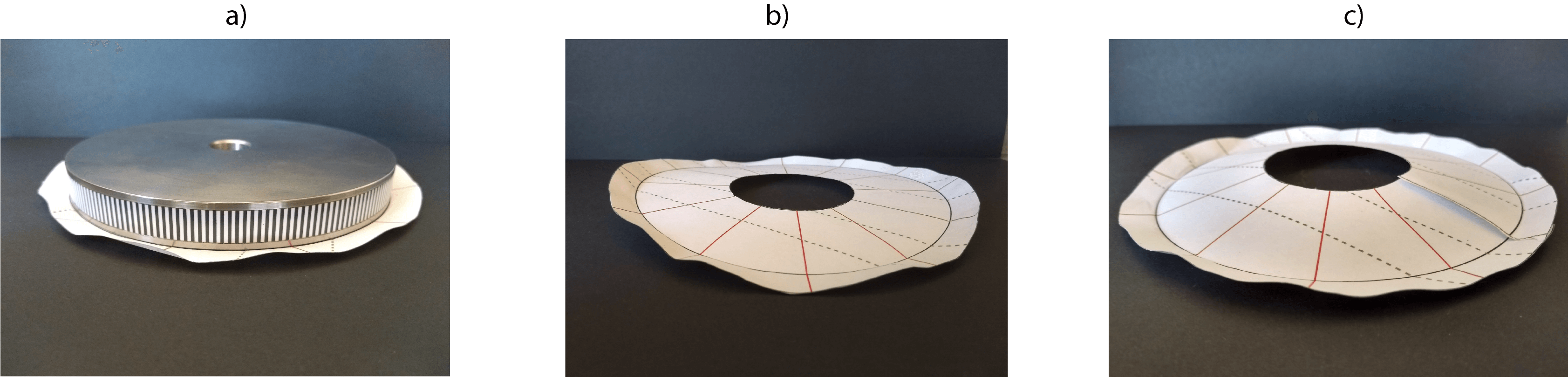}
\caption{Motivating example for the unfixed pivot section. a) Internal clamped rotation of a paper annulus leads to azimuthal ruffling. We crease the paper about the disk, to fix the angle $\alpha$. What happens when we remove the disk, rendering the pivot circle free to translate? b) The uncut annulus buckles up to form an $m=2$ saddle shape, exactly like the curved crease. c) A radially cut annulus instead overlaps to form a double overlapping c-cone (a 'bi-cone').  }
\label{fig:creaseProgression}
\end{figure}
\paragraph{}
We motivate this section using the following extension to the $\rho_p \neq 0$ clamped rotation experiment described above. We start with a paper annulus, its interior clamped down with a heavy metal disk, and the exterior rotated inwards by some angle $\alpha_{\rm out}$. We then intentionally crease the paper about the pivot, so that the paper deforms plastically and the rotation angle stays constant even after we stop actively rotating the paper. With the rotation angle fixed, what happens if we now remove the metal disk? (Figure \ref{fig:creaseProgression}) As expected for a circular creased annulus \cite{dias_geometric_2012,dias_non-linear_2014}, it responds elastically by buckling, which means that the wrinkled annulus had residual compressive strain. Moreover, the amplitude of buckling depends strongly on the width of the inner region. However, if we cut up the annulus radially before lifing the metal disk, then there is no buckling response; instead, the annulus curls in and overlaps. Recall that the clamping has two consequences for elastic strains -- i) it eliminates the interior of the pivot circle from the problem, and ii) it fixes the pivot circle at a constant radius $\rho_p$. Removing it thereby allows in-plane displacements, and the possibility of accessing lower-energy shapes that were previously forbidden, with the final shape determined by both the exterior and the interior of the pivot circle.

\begin{figure}[htb]
\centering
\includegraphics[width=\textwidth]{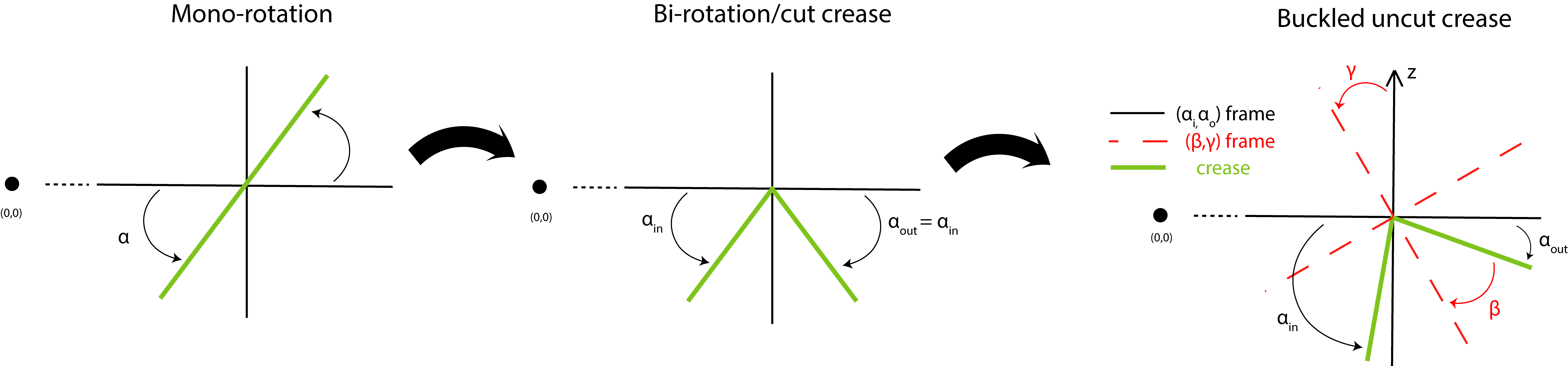}
\caption{Schematic of the succession of shapes discussed in section \ref{sec:creases}, interpolating between the conical solution and the curved crease. (Left) A mono-rotation -- specified by a single rotation angle $\alpha$ -- on a cut annulus, which creates a planar overlapping cone. (Centre) A bi-rotation, given by two rotation angles ($\alpha_{\rm in}, \alpha_{\rm out}$), on a cut annulus, which creates a planar bi-cone with constant $\alpha_{\rm in} = \alpha_{\rm out }$. (Right) A bi-rotation on an uncut annulus, which creates a buckled creased annulus with oscillating $\alpha_{\rm in}$ and $\alpha_{\rm out}$. In this case, it is useful to change from the  ($\alpha_{\rm in}, \alpha_{\rm out}$) rotation basis to the ($\beta, \gamma$) creaseline basis, as explained in the appendix.  }
\label{fig:crease}
\end{figure}

\paragraph{}
In this section, we investigate these different elastic responses using our framework of circles, using the conical solutions described above as a reference point. To this purpose, we consider three cases of \textit{unclamped} $\rho_p \neq 0$ rotations, in increasing order of complexity (Figure \ref{fig:crease}). First, we consider the unclamped mono-rotation of an annulus (or disk) cut up along a radial line (henceforth referred to as a 'cut annulus'). We find that this always gives rise to a regular overlapping c-cone, irrespective of the choice of pivot radius. We next consider a bi-rotation, necessarily about $\rho_p \neq 0$, in the same cut annulus. Finally, we consider the bi-rotation in a regular, uncut annulus. These last two systems correspond, respectively, to 'cut creases' and 'curved creases' \cite{dias_non-linear_2014}, and differ in their elastic response to the developed strains. The cut crease behaves exactly like two copies of the overlapping c-cone (hence termed a 'bi-cone'), while the full curved crease, because of its closed topology, must necessarily buckle in order to get rid of excess length. We show that this buckled shape, however, is not equivalent to two copies of the d-cone.

\subsection{Rotation of a cut annulus: the overlapping c-cone}

We perform the following experiment with a paper annulus. We make a radial cut in it, and rotate it inwards about its inner radius $R_{\rm in}$ by some angle $\alpha_c$. At first sight, this seems to be a rotation about a pivot circle ($\rho_p \neq 0$) that generates $\epsilon_{\theta \theta} <0$ according to (\ref{eq:rotationClampedStrain}), as discussed in the previous section. However, $\rho_p$ is no longer fixed by clamping, and this fact markedly changes the response to this deformation. The cut annulus now overlaps itself to get rid of excess length, while shrinking inwards at the same time, i.e. decreasing $\rho$ for the entire sheet. More surprisingly, the same thing happens if we rotate the annulus outwards about the outer radius $R_{\rm out}$, a deformation that should generate $\epsilon_{\theta \theta} >0$ according to (\ref{eq:rotationClampedStrain}), and hence no overlap at all! 
\paragraph{}
The only way this is possible is if the rotations, in the absence of clamping, are taking place about the central axis, $\rho_p =0$, thereby creating a cone! We can also see this in the physical shape of the rotated annuli (Figure \ref{fig:rotationShapes}b) -- they have constant angular overlap, and are unaffected by system length scales. Thus, the internal rotation about $R_{\rm in}$ makes a cone with its tip somewhere below $z=0$, while the external rotation about $R_{\rm out}$ makes a cone with the tip above $z=0$. In turn, the only way for an attempted rotation about $\rho_p \neq 0$ to become a rotation about $\rho_p =0$, is for the rotation to be accompanied by an extra displacement inwards, $u_\rho = -\rho_p$. As a result, the planar differential arc $dc$ located at polar coordinates $(r,\theta)$ retains its rotation axis $\hat e_{\theta}$, but translated inwards by an amount $-\rho_p$. 
\paragraph{}
What happens if we repeat this experiment with an uncut annulus? Given the good boundary conditions, $\alpha(\theta) \geq\alpha_c$, we get back the d-cone. The behaviour seen here is entirely consistent with the fact that conical solutions are the only zero-strain solutions possible under radial rotations. The clamping of the $\rho_p \neq 0$ rotations had suppressed these solutions, but removing it enabled the sheet to recover them again. However, as we will see below, changing the mono-rotation to a bi-rotation, it is possible to again suppress the conical solution, using not clamping but other constraints.

\subsection{The cut crease as two overlapping c-cones}
We now go back to the experiment described at the beginning of this section. With the annulus cut up, removing the metal disk frees up the pivot circle to displace radially, and as with the c-cone, the compressed circles overlap to get rid of excess length. Since the pivot circle is now a creaseline, it is easier to study this response by directly studying a planar cut annulus, creased along its midline. The question we ask is -- is this 'cut crease' equivalent to two independent overlapping c-cones, joined together along the creaseline? Experiments with paper creases suggest that the answer is clearly yes (Figure \ref{fig:rotationShapes}f). This is supported by three convincing arguments -- i) all the circles are planar, ii) the shape is insensitive to its inner and outer widths (i.e. scale-invariance), and iii) there is a constant angular overlap. 
\paragraph{}
Thus, the creaseline, for a cut crease, seems to act as only a pivot circle for engendering opposite-signed rotations on its two sides. The resulting shape may be termed a 'bi-cone', with the outer cone being rotated about its inner radius by angle $\alpha_{\rm out}$, and the inner cone being rotated about its outer radius by angle $\alpha_{\rm in}$. Moreover, since the axes of the two cones are parallel to the $z$-axis, we have $\alpha_{\rm out} = \alpha_{\rm in}$. Thus, ultimately, the cut crease is completely defined by a single angle, not two.

\subsection{The buckled curved crease}
What happens if we glue the two ends of this cut crease together, or equivalently, if we crease an uncut annulus? The resultant state will be one of azimuthal compression, and the only way to relieve this is through azimuthal buckling (Figure \ref{fig:intro}c and \ref{fig:rotationShapes}g). As before, we ask ourselves if this buckled shape behaves like two cones -- d-cones this time, because of the topology -- joined at the creaseline. The answer this time is clearly no, and evidence is provided by many sources. Firstly, experiments with paper creases clearly show that the shape, and the amplitude of buckling in particular, is highly sensitive to the inner and outer widths. Secondly, \cite{dias_geometric_2012} explicitly calculates the generators/directors of the surface in the limit of narrow creases, and shows that neither those on the inside nor on the outside are conical (i.e. converging to a point). Again in the narrow crease limit, where the creased annulus can be treated as a rod, \cite{mouthuy_overcurvature_2012} derive an explicit expression for the buckled creaseline which is clearly not a spherical curve, and hence not part of a cone.   
\paragraph{} 
However, at the same time, we see that the lowest buckling state is a pure saddle-shape with wavenumber $m=2$ -- a global buckling, as seen in the d-cone, and indicative of the absence of external constraints. Thus, we find that the buckled circular crease represents a novel solution in the category of radial rotations, clearly not conical ($\rho_p=0$) and yet not clamped ($\rho_p\neq 0$) either. The constraint here, in fact, is imposed not by an external clamping, but by the creaseline and its imposed dihedral angle, which constrains the surface directors to be non-radial in the buckled state \cite{dias_geometric_2012}. Thus here, we find the limitations of our framework, where it must act in complement to the crease-constrained generators in order to fully describe the surface. However, as shown in Appendix C, it is sufficient to predict the general form of the buckling.

\section{Summary and discussion}

In this paper, we have presented a unified geometric framework for studying the large-scale properties of planar thin sheet buckling under axisymmetric deformations. This framework exploits the axisymmetry of the loading to introduce macroscopic line elements -- radial lines and concentric circles -- that allow us to investigate even global deformations like radial rotations. Since the novelty of axisymmetric geometry lies in the curved boundaries, we concentrated on the circular line elements of the surface and their deformations (viz. the azimuthal strain $\epsilon_{\theta \theta}$), which gives rise to azimuthal buckling. We classified radial deformations into two broad categories -- in-plane deformations, which generate both radial and azimuthal strain, and out-of-plane deformations, which only generate azimuthal strain. Then, using a systematic classification scheme, and a geometric analysis of the buckled solutions, we recovered three azimuthally buckled solutions that have hitherto been unconnected -- azimuthal wrinkling, the d-cone and the circular curved crease. Along with these, we also recovered a class of azimuthal wrinkling resulting from clamped rotations about circular boundaries (seen, e.g., in cloths draped over tables), and predicted a class of axisymmetric radial buckling under radial in-plane compression that has not yet been studied, to the best of the author's knowledge, and is the subject of parallel research.

\paragraph{}
In particular, for the category of radial rotations, we saw that our framework acts as a complement to the traditional framework of directors used for isometric sheets. Indeed, for radial rotations about fixed pivots (both points and circles), the two frameworks coincide, since the directors coincide with the radial lines. On the other hand, for the buckled curved crease, created via radial rotation about an unfixed pivot, the two frameworks remain complementary; since our frameworks describes the role of surface elasticity, while the directors decribe the constraint imposed by the creaseline, we expect that used together, they will suffice to solve the curved creased problem for arbitrary widths. This will be explored in a future publication. 

\paragraph{}
Moreover, since our framework unifies azimuthal wrinkling and director-determined structures like the d-cone and curved crease, one might ask if and how the director framework is applicable to azimuthal wrinkling? The answer to this question is still open; while recent work has reported the emergence of an 'asymptotic isometry' in azimuthal wrinkling deep into the FT regime \cite{vella_wrinkling_2015,vella_indentation_2015}, these isometric sheets are clearly not developable, i.e. defined by generators. 


\paragraph{}
Although our framework exploits axisymmetry to introduce global line elements, the expressions (\ref{eq:azimuthalStrain}) and (\ref{eq:rotationClampedStrain}) for the azimuthal strain $\epsilon_{\theta \theta}$ are completely local. This strongly suggests that our global framework can be extended even to non-axisymmetric situations, i.e. where the deformations are applied along boundaries that are not circular. This would require one key modification -- the concentric circles must be modified to reflect the new boundary curve, and the radial distance $r$ (or $\rho$) in (\ref{eq:azimuthalStrain}) and (\ref{eq:rotationClampedStrain}), which is constant for a circle, must be replaced by the local radius of curvature. The other ingredients of our buckling analysis, viz. boundary conditions and external constraints, are global and thus remain unmodified. In the absence of axisymmetry, since there is no longer a global centre (or central axis), the conical solutions will no longer exist; but all the other buckled structures should still exist, with all their large-scale, geometric properties preserved. 

\begin{figure}[htb]
\centering
\includegraphics[width=0.8\textwidth]{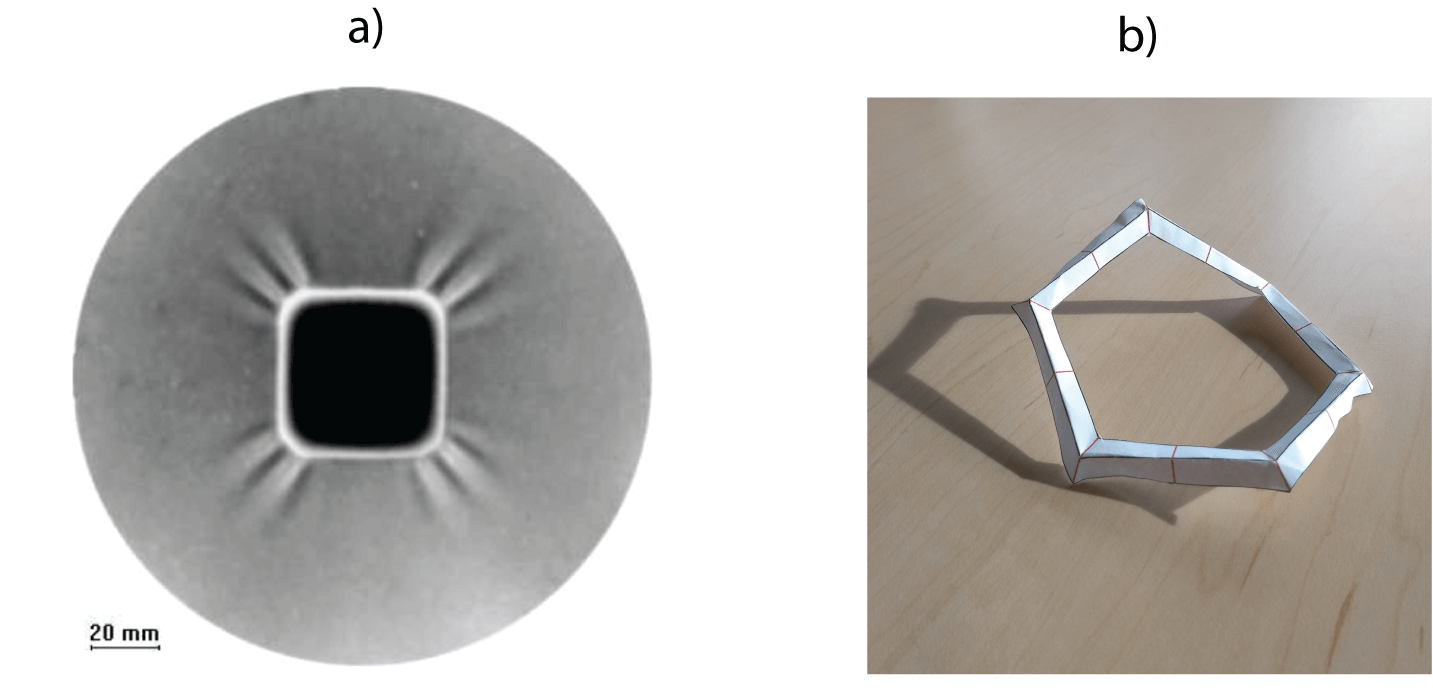}
\caption{Generalisation to non-axisymmetric geometries. We consider two extreme cases where the boundary curvature is localised at discrete points. a) The Lam\'e problem with a squarish inner boundary (from \cite{geminard_wrinkle_2004}) instead of a circular one, produces the same confined azimuthal wrinkling, but only around the points where the boundary curvature is localised. b) The $m=2$ buckling of a 'curved' crease with a hexagonal creaseline.}
\label{fig:discussion}
\end{figure}

\paragraph{}
This claim of generalisation, which will be properly tested in future work, is strongly supported by experimental results as well. For example, in Figure \ref{fig:discussion}, we show two egregious cases of arbitrarily curved deformation boundaries, where the curvature is almost or completely localised. The left image (taken from \cite{geminard_wrinkle_2004}) shows confined axial wrinkling under an inward pull, in an adaptation of the Lam\'e geometry to a rounded square; the right image shows a 'curved crease' with its characteristic $m=2$ buckling, but for a hexagonal creaseline. As can be seen, the large-scale properties of the buckling are the same as in the axisymmetric case. If our claim proves to be correct, then the geometric classification presented in this paper for axisymmetric deformations and buckling, with the exception of the conical solutions, should extend to non-axisymmetric cases as well, thus providing a general taxonomy of the shapes that can be created from a planar sheet through buckling under deformations acting along an arbitrarily curved boundary. 


\section{Acknowledgement}
The author is grateful to his thesis advisor, Thomas A. Witten, for enlightening discussions and extensive help with the manuscript. He would also like to thank Aaron Mowitz, Luka Pocivavsek and Menachem Stern for helpful comments, and Marc Downie for aid with the 3D figures.This work was performed as part of the author’s PhD research, and was supported partially by the National Science Foundation’s MRSEC program under award number DMR-1420709.

\section{Appendix A: Zeroes of $u_r$ under radial displacements}
As mentioned in the main text, since azimuthal strains generated by any radial displacement $u_r$ generate radial strains, and vice versa, the final self-consistent $u_r$ field in the sheet is determined by energy minimisation. The result, obtained by solving the plain strain equations (see, e.g., \cite{gould_introduction_2013}), is given by the general form
\begin{equation}
u_r (r) = A r + \frac{B}{r}.
\end{equation}
We are interested in the sign of $u_r$, and in particular, whether it changes sign within the sheet; thus, we must investigate the signs of $A$ and $B$. For boundary displacement loads $\Delta_i$ and $\Delta_o$at the inner and outer boundaries $R_i$ and $R_o$ resp. of an annulus, we obtain the following expressions for the constants:
\begin{equation}
A = \frac{\Delta_o R_o - \Delta_i R_i }{R_o^2 - R_i^2} \; , \; B = \frac{\Delta_i R_i R_o^2 - \Delta_o R_i^2 R_o }{R_o^2 - R_i^2} 
\end{equation}
To see whether $u_r$ changes sign, we must investigate the sign of the ratio. In particular, we can see that $u_r = 0 $ at the value $r_0 = \sqrt{-B/A}$; thus, for this zero to exist, we need $A/B <0$. The general expression is given by
\begin{align}
    \frac{A}{B} &=\frac{\Delta_o R_o - \Delta_i R_i}{\Delta_i R_i R_o^2 - \Delta_o R_i^2 R_o} \\
    					&= \frac{1}{R_i R_o} \frac{\frac{\Delta_o}{\Delta_i} - \frac{R_i}{R_o}}{1 -\frac{\Delta_o}{\Delta_i} \frac{R_i}{R_o}}.
\end{align}
For the simple case of $\Delta_i = \Delta_o$, we have $A/B = \frac{1}{R_i R_0} >0 $. Thus, $u_r$ never changes sign; this proves the assertion made in the main text that if the inner and outer boundary circles are both pulled in (out) equally, then \textit{all} the circles inside the material are pulled in (out). 
\paragraph{}
Again, if $\Delta_o/\Delta_i <0$, i.e. if both edges are pulled or pushed, we see that $A/B <0$, so that $r_0$ exists. This shows that in order for the Lam\'e wrinkles to be confined, it suffices to have an arbitrarily small external displacment, $\Delta_o >0$ .
\paragraph{}
Lastly, for a generic load,  $A/B >0$ if and only if the loads obey the inequality $0 < \frac{R_i}{R_o} < \frac{\Delta_i}{\Delta_o}$. Thus, $u_r$ changes sign iff $\frac{\Delta_i}{\Delta_o} < \frac{R_i}{R_o}$. Thus, interestingly, there is a range of values of ($\Delta_i$, $\Delta_o$) where they can have the same sign and still produce a change in the sign of $u_r$.

\section{Appendix B: Properties of conical (i.e. $\rho_p =0$) solutions}
We start by restating eq. (\ref{eq:coneStrain}), which shows that the circles are all equally compressed under a radial rotation of angle $\alpha$: 
\begin{equation}
\epsilon_{\theta \theta} = \cos \alpha -1 = \rm const.,
\end{equation} which means such a deformation is independent of any system length scale. As a result, \textit{all the circles must buckle similarly}, with only the buckling amplitude scaling with $r$ since the length of each circle scales with $r$. This scale-invariance is the fundamental property of conical solutions, from which the following properties follow.

\subsubsection{Cones are the unique strain-free ($\epsilon_{\theta \theta} = \epsilon_{rr}=0$) solutions under radial rotations}
Consider the ensemble of axisymmetric circles comprising the surface. In the unstrained planar state, these are all related by a linear radial scaling, so that defining one circle defines all the others. Equivalently, axisymmetry implies that the two-dimensional planar sheet is defined by a point Gaussian charge, here the angular extent $\Delta\Phi =2 \pi$. Under an axisymmetric out-of-plane deformation, the only way for this ensemble of circles to live together without stretching the radial lines is for the linear radial scaling -- i.e. for the point symmetry -- to be maintained also in three dimensions. For this, the concentric circles must be replaced by concentric spheres; in which case again, the entire two-dimensional surface can be obtained by radial translation of some curve lying on the sphere. By definition, this is exactly a conical solution.

\subsubsection{A conical circle of planar radius $r$ stays on the $r$-sphere }
 As stated, to describe the buckling of the entire surface, it suffices to consider any one circle. If this compressed circle is allowed to buckle and regain its full length, then it means that each differential arc must regain its original length $dc = r d\theta$. In the buckled state, let it be at distance $r'$ from the origin, and subtending angle $d\theta'$; thus, we have $dc = r' d\theta' = r d\theta$. But since the radial directors bordering the arc are unstrained, it means $r'=r$ (and also $d\theta'=d\theta$). Thus, the differential arc must remain on the $r$-sphere in the buckled state, and hence, so must the entire buckled circle! This is a strong statement, since a spherical curve obeys a local constraint relating the curvature $\kappa$ and the torsion $\tau$ \cite{struik_lectures_1988}. As a result, the curve can be described using only a single parameter -- either $\kappa$ or $\tau$ -- instead of requiring both (c.f. the fundemental theorem of space curves). This property was fully exploited in the original d-cone and e-cone papers \cite{cerda_dcones_1998, cerda_dcones_2005, muller_conical_2008}).

\section{Appendix C: Buckling of the curved crease in our framework }
A full study of the curved crease problem (of arbitrary width) requires its own space. Here, we limit ourselves to framing the problem through the lens of our framework (viz., the rotations $\alpha_{\rm in}$ and $\alpha_{\rm out}$). Consider the curved crease made at radius $\rho_{\rm crease}$. For a crease, it is useful to change the angular basis from $(\alpha_{\rm in},\alpha_{\rm out}) \to (\beta,\gamma)$ (Figure \ref{fig:crease}):
\begin{align} \label{eq:coneVariables}
\begin{split}
    \beta &= \pi/2 - (\alpha_{\rm in}+\alpha_{\rm out})/2 \\
    \gamma &= (\alpha_{\rm out}-\alpha_{\rm in})/2
 \end{split}
\end{align}
where $2\beta$ is the dihedral angle of the crease, and $\gamma$ is the twist/torsion angle giving angular deviation of the bisector of the dihedral angle from $\hat z$ (note that we have changed the sign of $\alpha_{\rm in}$ with respect to the usual definition). $\beta$ can also be thought of as the mean rotation angle, and $\gamma$ as the difference in rotational load between the two sides. The advantage of using the ($\beta,\gamma$) basis is that the boundary condition imposed by a crease is defined in terms of its dihedral angle, $\beta(\theta)$. For an axisymmetric loading, this must be constant $\beta (\theta)=\beta_0$, and we can take it to be constant even in the buckled state without changing the large-scale characteristics (as seen in\cite{mouthuy_overcurvature_2012}). As a result, the only angular freedom lies in $\gamma (\theta)$. Moreover, since we are explicitly looking for non-conical shapes, the rotation pivot radius $\rho_p (\theta)$ gives us an extra translational variable, which for now, we take to be independent from $\gamma$. Thus, these two variables serve to completely describe the creased annulus (the amplitude of buckling is given by the torsion angle, $\gamma (\theta)$, by a geometrical relation). 
\paragraph{}
In this basis, the cut crease is given by the constant solution: $(\gamma, \rho_p) = (0, \rho_{\rm crease} \sin \beta_0$). The uncut crease will give a buckled solution $\gamma(\theta)$, obeying periodic boundary conditions. The lowest energy solution has the mode number $m=2$ (consistent with the four-vertex theorem). Inverting (\ref{eq:coneVariables}), this means that the buckled solution will have $\alpha_{\rm in} (\theta)$ and $\alpha_{\rm out} (\theta)$ oscillating in phase about their mean value $(\pi/2-\beta_0) $:
\begin{align}
\begin{split}
    \alpha_{\rm in}(\theta) &= (\pi/2-\beta_0) -\gamma(\theta) \\
    \alpha_{\rm out}(\theta) &= (\pi/2-\beta_0) +\gamma(\theta) 
 \end{split}
\end{align}
Physically, this looks like the swaying of two legs in-phase about a central bar (see Figure \ref{fig:intro}c). This solution was calculated analytically in the limit of the narrow crease \cite{dias_geometric_2012,dias_non-linear_2014}. 

\bibliographystyle{unsrt}
\bibliography{bibliography.bib}

\end{document}